\documentclass[letterpaper, 10 pt, conference]{ieeeconf}  

\usepackage{graphicx}
\usepackage{tabularx,booktabs}
\usepackage{color}
\usepackage{multicol}
\usepackage{amsmath,amssymb}
\usepackage{amsfonts}
\usepackage{textcomp}
\usepackage{psfrag}
\usepackage{multimedia}
\usepackage{fancybox}
\usepackage{algorithm}
\usepackage{algorithmic}
\usepackage{varioref}
\usepackage{import}
\usepackage{framed,xcolor}
\usepackage{color}
\usepackage{float}
\usepackage{etoolbox}
\usepackage{soul}

\IEEEoverridecommandlockouts                              

\definecolor{seb}{rgb}{0.8,1,0.8}

\definecolor{hos}{rgb}{0.91,0.84,0.42}

\newcommand{\vect}[1]{\ensuremath{\boldsymbol{\mathrm{#1}}}}

\DeclareMathOperator*{\argmin}{arg\,min}
\title{\LARGE \bf
	Reinforcement Learning based on MPC/MHE for Unmodeled and Partially Observable Dynamics
}

\author{Hossein Nejatbakhsh Esfahani, Arash Bahari Kordabad, S\'ebastien Gros
	\thanks{Authors are with Department of Engineering Cybernetics, Norwegian
		University of Science and Technology (NTNU), Trondheim, Norway.
		{\tt\small \{hossein.n.esfahani, arash.b.kordabad, sebastien.gros\}@ntnu.no%
		}
}}

\begin{document}
	\maketitle
	\begin{abstract}
		This paper proposes an observer-based framework for solving Partially Observable Markov Decision Processes (POMDPs) when an accurate model is not available. We first propose to use a Moving Horizon Estimation-Model Predictive Control (MHE-MPC) scheme in order to provide a policy for the POMDP problem, where the full state of the real process is not measured and necessarily known. We propose to parameterize both MPC and MHE formulations, where certain adjustable parameters are regarded for tuning the policy. In this paper, for the sake of tackling the unmodeled and partially observable dynamics, we leverage the Reinforcement Learning (RL) to tune the parameters of MPC and MHE schemes jointly, with the closed-loop performance of the policy as a goal rather than model fitting or the MHE performance. Illustrations show that the proposed approach can effectively increase the performance of close-loop control of systems formulated as POMDPs.
	\end{abstract}
	
	\section{Introduction}
	Reinforcement Learning (RL) is a powerful tool for solving Markov Decision Processes (MDP) problems \cite{sutton}. RL methods often use Deep Neural Network (DNN) to approximate either the optimal policy underlying the MDP directly or the action-value function from which the optimal policy can be indirectly extracted.
	\par
	Recent publications are discussing RL for POMDPs. A neural network-based computation of belief states (posterior distributions over states) was proposed to aggregate historical information needed to estimate a belief state \cite{belief1,belief2}. An RL algorithm tailored to POMDPs was proposed in \cite{thesis1} that incorporated spectral parameter estimation within an exploration-exploitation strategy. A data-driven algorithm based on approximate Dynamic Programming (ADP) was proposed in \cite{r6} to stabilize a plant with partially observable dynamics. The authors used an action-dependent heuristic dynamic programming (ADHDP) algorithm including two neural networks as an actor-critic (AC) method to estimate both the unmeasured state and the performance index. The proposed ADP-based approach in \cite{r5} is similar to classic RL algorithms but requires only measurements of the input/output data and not of the full system state.
	\par
	In \cite{r1}, a neural network-based actor-critic structure was proposed to approximate the control policies where a full system state is not accessible. In~\cite{ELearning}, a fuzzy neural network was used to find the local optimal policy. A recurrent neural network (RNN) was proposed in \cite{r7} to learn and infer the true state observations from the noisy and correlated observations in a POMDP. Most of the proposed RL-based control techniques in the above literature are based on DNN-based approximators.
	
	\par 
	Model predictive control (MPC) is a popular and widely used practical approach to optimal control. MPC is often selected for its capability to handle both input and state constraints \cite{rawling}. At each time instant, MPC calculates the input and corresponding state sequence minimizing a cost function while satisfying the constraints over a given prediction horizon. The first input is applied, and the optimal solution is recalculated at the next time instant based on the latest state of the system.
	\par 
	In many practical applications, some states of the plant are estimated using an observer since they can not be directly measured, and the plant is possibly not fully observable. The Moving Horizon Estimation (MHE) is a well-known model-based observer in order to estimate the states of processes. In this paper, we use this type of observer as a natural choice for the MPC scheme \cite {MS}. 
	
	\par 
	Recently, the integration of machine learning in model predictive control has been presented, with the aim of learning the model of the system, the cost function or even the control law directly \cite{sergio1,sergio2}. These approaches are based on DNN-based approximation. The direct combination of RL and MPC has been investigated in \cite{rl1,rl2,rl3}. It is shown that a single MPC scheme can capture the optimal value function, action-value function, and policy of an MDP, even if the MPC model is inaccurate, hence providing a valid and generic function approximator for RL. The applications of this new MPC-based RL framework have been recently presented in \cite{Arash2021MPC,Arash2021ACC}.
	\par
	However, these approaches assume that the state of the process is known and can be fully measured. For many applications this assumption is not fulfilled. To address this issue, this paper proposes to use a state observer such as an MHE combined with the MPC scheme to build a policy based on the historic of the available measurements rather than on the full state of the system. MHE delivers state estimations by fitting the process model trajectory to past measurements obtained on the real system. We adopt an MPC-based Q-learning algorithm to tune the parameters included in the MHE-MPC scheme for the closed-loop performance of the resulting policy.
	
	\par 
	This paper is organised as follows. In Section \ref{sec:Preliminaries and problem formulation}, some background material is given. Then the MPC and MHE schemes are detailed. The implementation of the Q-learning algorithm for tuning both the MPC and MHE schemes together is detailed in Section \ref{sec:MPC/MHE-based RL}. An illustrative example is proposed in Section \ref{sec:Numerical Example}. Finally, conclusions and future work are given in Section \ref{sec:Conclusion}.
	
	\section{Preliminaries and problem formulation} \label{sec:Preliminaries and problem formulation}
	In the context of reinforcement learning a partially observable real plant is described by a discrete POMDP having (possibly) stochastic state transitions as follows:
	\begin{subequations}
		\label{eq:plant}
		\begin{gather}
			\vect x_{k+1}=f^{\text{plant}}(\vect x_k,\vect u_k,\vect\zeta) \label{eq::model}\\
			\vect y_k=h(\vect x_k,\vect\eta)\label{eq::out}
		\end{gather}
	\end{subequations}
    where the full state $\vect x_k$ is not measurable or not even known and $\vect x_{k+1}$ is the next plant state vector under stochastic transition with some random disturbances $\vect \zeta$. The model outputs $\vect y$ are measured on the real system and delivered by the output function $h$  associated with some random measurement noises $\eta$. We present next the MHE and MPC schemes and how they can be used to create the action-value function approximation required in Q-learning.
	
	\subsection{Parameterized MHE Formulation}
	For a POMDP, the measurements available from the real process at a given time instant do not constitute a Markov state. As a result, the full history of the measurements becomes possibly relevant to the optimal policy. However, building a policy based on the complete measurement history to solve the POMDPs is not realistic. The RL community either considers a limited sequence of past observations as a sufficient history or estimates a belief state using a recurrent neural network. In this paper, we propose a more structured solution to address this issue, by using MHE as a model-based approach to build a state from the measurement history. The complete measurement history is then transformed into a (possibly small) model state that is compatible with the selected policy.
	\par  The MHE-based observer at the physical time $k$ can be stated as the following Nonlinear Least-Squares problem:
	\begin{subequations}\label{eq:MHE}
		\begin{align}
			&\left \{ \hat{\vect x}_{k-N_{\text{MHE}},\ldots,k},\hat{\vect u}_{k-N_{\text{MHE}},\ldots,k-1} \right \}\nonumber\\
			&=\argmin_{\vect x,\vect u}\left \| \vect x_{k-N_{\text{MHE}}}-\tilde{\vect x}_{k-N_{\text{MHE}}} \right \|_{A^\theta_{r}}^2  \nonumber\\
			&\quad+ \sum_{i=k-N_{\text{MHE}}}^{k}\left \|\bar{\vect y}_i-\vect y(\vect x_i) \right \|_{Q^\theta_{E}}^2 +\phi_\theta(\vect x_i)\nonumber\\
			&\quad+\sum_{i=k-N_{\text{MHE}}}^{k-1}\left \| \vect u_i-\bar{\vect u}_i \right \|_{R^\theta_{E}}^2 +\phi_\theta(\vect u_i) \label{eq:arrival}\\ 
			&\quad\mathrm{s.t.}\quad \vect x_{i+1}=f_\theta^{\text{MHE}}(\vect x_i,\vect u_i)
		\end{align}
	\end{subequations}
	where $k$ is the current time instant, $i$ is the time instant along the estimation horizon window. $\bar{\vect y}_i,\bar{\vect u}_i$ are the measurements available at the physical time $k$ while their corresponding values obtained from the MHE model are $\vect y(\vect x_i),\vect u_i$, respectively. Let us to consider the mismatch between the model (observer) and the real plant measurements is explainable by normal centered output noise, then matrices $Q^\theta_E$ and $R^\theta_E$ are the inverse of the covariance matrices associated to these noises on the plant output and control input measurements, respectively. The first term in \eqref{eq:arrival} is an arrival cost weighted with matrix $A^\theta_r$, which aims at approximating the information prior to $k-N_{\text{MHE}}$, where $\tilde{\vect x}$ is the available estimation for the state at time $k-N_{\text{MHE}}$. In practice, since the MHE fitting error is not only coming from some normal centered output noise but also model error, more intricate noise, and possibly unmodelled dynamics, it is very difficult to decide what symmetric positive semi-definite weighting matrices  $Q^\theta_E$, $R^\theta_E$, $A^\theta_{r}$  ought to be used to obtain the best closed-loop performance. To address this issue, we propose to adjust them using the RL algorithm. Moreover, as the Least-Squares cost as a choice of penalty in the MHE are not necessarily sufficient, we introduce a cost modification $\phi_\theta$ tuned by RL. Note that we consider a gradient form of the cost modification in this paper $\phi_\theta(\vect x_i)=f_1^\top\vect x_i$ and $\phi_\theta(\vect u_i)=f_2^\top\vect u_i$, where $f_1$ and $f_2$ are labeled as RL parameters $\theta$.

	\subsection{Parameterized MPC Formulation}
	\par
	In this work we will consider the MPC scheme as a value function approximator that can be formulated as:
	\begin{subequations}\label{eq:V}
		\begin{align}
			V_{\theta }(\vect x_k)=&\min_{\vect x,\vect u,\vect \sigma }\quad \gamma^{k+N_{\text{MPC}}}\left(V^f_{\theta}(\vect x_{k+N_{\text{MPC}}})+\vect w_{f}^\top\vect\sigma_{k+N_{\text{MPC}}}\right) \nonumber\\
			&\quad +\sum_{i=k}^{k+N_{\text{MPC}}-1}\gamma^{i}\left(l_{\theta}(\vect x_{i},\vect u_{i})+\vect w^\top\vect \sigma_{i}\right)\label{eq:v1}\\
			\mathrm{s.t.}&\quad \vect x_{i+1}=f_{\theta}^{\text{MPC}}(\vect x_i,\vect u_i),\label{eq:mpc_bias}\\
			&\quad\vect x_k=\hat{\vect x}_k, \label{eq:v2}\\
			&\quad \vect g(\vect u_{i})\leq 0,\\
			&\quad \vect h_{\theta}(\vect x_{i},\vect u_i)\leq \vect \sigma_{i},\quad \vect h_{\theta}^{f}(\vect x_{k+N_{\text{MPC}}})\leq \vect \sigma_{k+N_{\text{MPC}}}  \\
			&\quad \vect\sigma_{k,\ldots,k+N_{\text{MPC}}-1} \geq 0 \label{eq:violation}
		\end{align}
	\end{subequations}
	
	\par 
	
	where $l_\theta$ is the stage cost, $V^f_\theta$ the terminal cost, $f_{\theta}^{\text{MPC}}$ the MPC model (possibly but not necessarily different from the MHE model), $\vect h_\theta$ the mixed constraints, $\vect g$ the pure input constraints, $\vect h_{\theta}^{f}$ the terminal constraints. The MPC initial conditions in \eqref{eq:v2} are delivered by MHE scheme at the current time instant $k$.	
	In many real processes, there are uncertainties and disturbances that may cause an MPC scheme to become infeasible. Therefore, an $\ell_1$ relaxation of the mixed constraints \eqref{eq:violation} is introduced. An exact penalty is imposed on the corresponding slack variables $\vect \sigma_k$ with large enough weights $\vect w,\vect w_f$ such that the trajectories predicted by the MPC scheme will respect the constraints. All elements in the above MPC scheme are parameterized by $\theta$, which will be adjusted by RL, as detailed in \cite{rl1}. 
	
	\par
	Let us consider the policy at the physical current time $k$ as:
	\begin{gather}\label{eq:Policy}
		\pi_{\theta}(\vect x_k)=\vect u_k^*
	\end{gather}
	where, $\vect u_k^{\star}$ is the first element of the input sequence $\vect u_k^{\star},\cdots,\vect u_{k+N_{\text{MPC}}-1}^{\star}$ solution of \eqref{eq:V}. We next consider this optimal policy delivered by the MPC scheme as an action $\vect a_k$ in the context of reinforcement learning where, it is selected according to the above policy with the possible addition of exploratory moves \cite{sutton}.
	Then, an action-value function approximation $Q_{\theta}$ can be formulated as:

	\begin{subequations}\label{eq:Q}
		\begin{align}
			Q_{\theta}(\vect x_k,\vect u_k)=\min_{\vect x,\vect u,\vect \sigma}&\quad \eqref{eq:v1}\\
			\mathrm{s.t.}&\quad \eqref{eq:v2}-\eqref{eq:violation}\\
			&\quad \vect u_k=\vect a_k
		\end{align}
	\end{subequations}
	
	 Note that the proposed approximations \eqref{eq:V}-\eqref{eq:Q} satisfies the fundamental equalities underlying the Bellman equations \cite{bertsekas}:
	\begin{gather}
		\pi_{\theta}(\vect x_k)=\arg\min_{\vect u} Q_{\theta}(\vect x_k,\vect u_k),\quad V_{\theta}(\vect x_k)=\min_{\vect u} Q_{\theta}(\vect x_k,\vect u_k) \label{bell}
	\end{gather}

	\section{MPC/MHE-based RL} \label{sec:MPC/MHE-based RL}
	In this section, we present the algorithmic details needed to implement a classic Q-learning algorithm on the combination of MPC/MHE schemes.
	\subsection{Q-Learning for MPC/MHE}
	A classical off-policy Q-Learning algorithm is based on the temporal-difference learning procedure \cite{sutton} in which the updating rule for the RL parameters can be expressed as follows:

	\begin{subequations}\label{eq:QL}
		\begin{align}
			\delta_k =L(\vect x_k,\vect u_k)&+\gamma V_{\theta}(\vect x_{k+1})-Q_{\theta}(\vect x_k,\vect u_k)\label{eq:delta},\\
			\theta \leftarrow \theta&+\alpha \delta_k \nabla_{\theta} Q_{\theta}(\vect x_k,\vect u_k)
		\end{align}
	\end{subequations}
    where scalar $\alpha>0$ is a step size, $0< \gamma\leq 1$ is a discount factor and $\delta_k$ is the Temporal Difference (TD) error at the physical time $k$. In the above TD learning algorithm, a baseline stage cost $L(\vect x_k,\vect u_k)$ (reward in the context of RL) is defined as a function of state-action pair in order to provide an evaluation signal. Indeed, the baseline cost affects the agent behavior and control policy via RL parameter updating, where the TD error is appeared.
	
	\par 
	The gradient of function $Q_\theta$ needed in \eqref{eq:QL} requires one to compute the sensitivities of the optimal value of Nonlinear Programming (NLP) \eqref{eq:Q}. This sensitivity ought to be computed with care since the RL parameters $\theta$ impact $Q_\theta$ both directly via the MPC scheme and indirectly via the MHE scheme, by modifying the state estimation $\hat{\vect x}_k$ at the physical current time instant $k$ that enters as an initial condition $\vect x_k=\hat{\vect x}_k$ in the MPC scheme. The gradient $\nabla_{\theta} Q_{\theta}$ associated to the proposed MPC/MHE scheme is given by the following total derivative:
	\begin{gather} \label{eq:sens}
		\frac{dQ_\theta}{d\theta}=\frac{\partial Q_\theta}{\partial \theta}+\frac{\partial Q_\theta}{\partial \vect x_k}\frac{\partial \hat{\vect x}_k}{\partial \theta},\quad \vect x_k=\hat{\vect x}_k
	\end{gather}
	We detail next how to compute the above sensitivities. 
	
	\subsection{Sensitivities of the MPC/MHE scheme}
	Let us define the Lagrange functions $\mathcal{L}_\theta,\hat{\mathcal{L}_\theta}$ associated to the MPC and MHE problems \eqref{eq:MHE}, \eqref{eq:Q} as follows:
	\begin{gather}
		\mathcal{L}_{\theta}=\Phi_{\theta}+\vect \lambda^\top G_{\theta}+\vect \mu^\top H_{\theta}\\
		\hat{\mathcal{L}}_{\theta}=\hat{\Phi}_{\theta}+\hat{\vect \lambda}^\top \hat{G}_{\theta}
	\end{gather}
	
	where $H_{\theta}$ gathers the inequality constraints of \eqref{eq:Q} and $\Phi_\theta,\hat{\Phi}_\theta$ are the costs of the MPC and MHE optimization problems, respectively.  Variables $\vect \lambda, \hat{\vect \lambda}$ are the Lagrange multipliers associated to the equality constraints $G_\theta,\hat{G}_\theta$ of the MPC and MHE, respectively. Variables $\vect\mu$ are the Lagrange multipliers associated to the inequality constraints of the MPC scheme. Let us label the primal variables as $\vect p=\left\{\vect X,\vect U\right\}$ and $\hat{\vect p}=\left\{\hat{\vect X},\hat{\vect U}\right\}$ for the MPC and MHE, respectively. The primal-dual variables of the MPC and MHE schemes will be labeled as $\vect z=\left\{\vect p,\vect \lambda,\vect\mu\right\}$ and $\hat{\vect z}=\left\{\hat{\vect p},\hat{\vect \lambda}\right\}$, respectively.
	
	The sensitivities of the MPC scheme \eqref{eq:Q} required in \eqref{eq:sens} can be obtained by the sensitivity analysis detailed in \cite{sensAnalysis} as follows:	
	\begin{gather}\label{eq:sens_mpc}
		\frac{\partial Q_\theta}{\partial \theta}=\frac{\partial \mathcal{L}_\theta(\vect x_k,\vect z^\star)}{\partial \theta},\quad 
		\frac{\partial Q_\theta}{\partial \vect x_k}=\frac{\partial \mathcal{L}_\theta(\vect x_k,\vect z^\star)}{\partial \vect x_k}
	\end{gather}		
	where $\vect z^\star$ is the primal-dual solution vector of \eqref{eq:Q}.
	
	\par 
	 The sensitivity $\frac{\partial \hat{\vect x}_k}{\partial \theta}$ associated to the MHE scheme can be obtained via using the Implicit Function Theorem (IFT) on the Karush Kuhn Tucker (KKT) conditions underlying the parametric NLP. Assuming that Linear Independence Constraint Qualification (LICQ) and Second Order Sufficient Condition (SOSC) hold \cite{nlp} at $\hat{\vect z}^\star$, then, the following holds:  		
	
	\begin{align}
		\label{eq:SensMHE}
		\frac{\partial \hat{\vect z}^\star}{\partial \theta}=-\frac{\partial R_\theta}{\partial \hat{\vect z}}^{-1}\frac{\partial R_\theta}{\partial \theta}
	\end{align}
	where 
	\begin{align}
		R_{\theta}=\begin{bmatrix}
			\nabla_{\hat {\vect p}}\hat{\mathcal{L}}_{\theta}\\\hat{G}_{\theta} 
		\end{bmatrix}
	\end{align}
	are the KKT conditions associated to the MHE scheme \eqref{eq:MHE}.
	As $\hat{\vect x}_k$ is part of $\hat{\vect z}^\star$, the sensitivity of the MHE solution  $\frac{\partial \hat{\vect x}_k}{\partial \theta}$ required in \eqref{eq:sens} can be extracted from matrix $\frac{\partial \hat{\vect z}^\star}{\partial \theta}$.
	
	\subsection{Constrained RL steps}
	The adjustable weighting matrices in the proposed parameterization of both MPC and MHE in \eqref{eq:MHE}, \eqref{eq:V} and \eqref{eq:Q} are tuned using Q-learning. As a requirement, the weighting matrices $Q^\theta_E$, $R^\theta_E$, $A^\theta_{r}$ must be positive semidefinite. However, the RL steps delivered by Q-learning do not necessarily respect this requirement, and we need to enforce it via constraints on the RL steps throughout the learning process. To address this requirement, we formulate a Semi-Definite Program (SDP) as a least squares optimization problem:
	\begin{subequations}\label{eq:sdp}
		\begin{align}
			\min_{\Delta\theta}&\quad\frac{1}{2}\left \| \Delta \theta \right \|^2-\alpha\delta_k\nabla_{\theta}Q_{\theta} (\vect x_k,\vect u_k)^\top\Delta\theta\\
			\mathrm{s.t.}&\quad Q^\theta_E({\theta + \Delta \theta})\geq 0 \\
			&\quad R^\theta_E({\theta + \Delta \theta})\geq 0\\
			&\quad A^\theta_{r}({\theta + \Delta \theta})\geq 0
		\end{align}
	\end{subequations}
	where we assume that the weighting matrices $Q^\theta_E,R^\theta_E,A^\theta_{r}$ are linear functions of $\theta$. Then, these matrices are updated in each time instant due to updating $\Delta \theta$, which is a solution of the above SDP scheme. The proposed learning process is described in the Alg. 1.
	\begin{algorithm}
	\caption{(MPC+MHE)-Based RL}
	\begin{algorithmic}\label{algorithm}
		\REQUIRE $\alpha,tol>0, \theta=\theta_0, \vect x_0, \vect u_0$
		\WHILE{Iter}
		\STATE 1. Measure output $\vect{y}_k$ from \eqref{eq::out} at current time $k$
		\STATE 2. Obtain $\hat{\vect x}_k$,$\frac{\partial \hat{\vect x}_k}{\partial \theta}$ from \eqref{eq:MHE} and \eqref{eq:SensMHE}
		\STATE 3. Obtain $\pi_{\theta}(\vect x_k)$, $V_{\theta }(\vect x_k)$ from \eqref{eq:V}
		\STATE 4. Exploration: $\vect u_{k}=\pi_{\theta}(\vect x_k)+d$, $d\sim \mathcal{N}\left(\mu,\sigma^2\right)$
		\STATE 5. Obtain $Q_\theta$, $\frac{\partial Q_\theta}{\partial \theta}$ from \eqref{eq:Q}, \eqref{eq:sens_mpc}
		\STATE 6. Assemble $\frac{dQ_\theta}{d\theta}$ from \eqref{eq:sens}
		\STATE 7. Apply $\vect u_k$ to the real plant: \eqref{eq::model}
		\STATE 8. Evaluate baseline $L(\vect x_k,\vect u_k)$
		\STATE 9. RL update:
		\STATE \quad- Obtain $\Delta\theta$ from \eqref{eq:sdp}
		\STATE \quad- $ \theta \leftarrow \theta+\Delta\theta$
		\STATE $k \leftarrow k+1$
		\ENDWHILE
	\end{algorithmic}
\end{algorithm}
	\section{Numerical Example}\label{sec:Numerical Example}
\begin{figure}[htbp!]
		\centering
		\includegraphics[width=.8\linewidth]{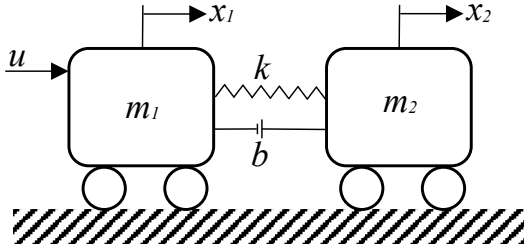}
		\caption{Two-Mass-Spring-Damper} 
		\label{POMDP}
	\end{figure}

	In this section, we illustrate the performance of the proposed MPC/MHE-based RL scheme, which is tested on a constrained two-mass-spring-damper system shown in Figure~\ref{POMDP}, for which the MPC/MHE model ignores some of the dynamics.
	\par 
	The control input acts on mass 1, and the position of mass 2 is measured. Let us consider $m_1=0.8$ kg, $m_2=0.5$ kg, $k=25$ $\frac{N}{m}$, $b=3$ $\frac{Ns}{m}$ and define the plant dynamics as:
	\begin{gather}
		\begin{bmatrix}
			\dot{x}_1\\ 
			\dot{x}_2\\ 
			\dot{x}_3\\ 
			\dot{x}_4
		\end{bmatrix}=\begin{bmatrix}
			\phantom{-}0 &\phantom{-}0  &\phantom{-}1  &\phantom{-}0 \\ 
			\phantom{-}0 &\phantom{-}0  &\phantom{-}0  &\phantom{-}1 \\ 
			-\frac{k}{m_1} & \phantom{-}\frac{k}{m_1} & -\frac{b}{m_1} &\phantom{-} \frac{b}{m_1}\\ 
			\phantom{-}\frac{k}{m_2}& -\frac{k}{m_2} & \phantom{-}\frac{b}{m_2} & -\frac{b}{m_2}
		\end{bmatrix}\begin{bmatrix}
			x_1\\ 
			x_2\\ 
			x_3\\ 
			x_4
		\end{bmatrix}+\begin{bmatrix}
			0\\ 
			0\\ 
			\frac{1}{m_1}\\ 
			0
		\end{bmatrix}u\nonumber
	\end{gather}
	where $x_1,x_2$ are positions of masses 1, 2, respectively, and $x_3,x_4$ are corresponding velocities. Variable $u$ is the control input applied to the first mass. 
	In this simulation, we propose to formulate the MPC/MHE scheme based on a partially observable model. More precisely, the adopted MPC scheme is presented based on a 2-states model, capturing only the position and velocity of mass 1. The MHE scheme is based on the same model, but is fed as measurements the position of the mass 2. Let $f_\theta^{\text{MPC}}$ in \eqref{eq:mpc_bias} be a partially observable and inaccurate MPC model as follows:
	\begin{gather}
		\begin{bmatrix}
			\dot{x}_1\\ 
			\dot{x}_3
		\end{bmatrix}=\left ( \begin{bmatrix}
			0 &1 \\ 
			0&0 
		\end{bmatrix}+A^{\text{bias}}
		\right )\begin{bmatrix}
			x_1\\ 
			x_3
		\end{bmatrix}+\left ( \begin{bmatrix}
			0\\ 
			\frac{1}{m_1+m_2}
		\end{bmatrix}+B^{\text{bias}} \right )u
	\end{gather}
    where $A^{\text{bias}}, B^{\text{bias}}$ are adjusted as model bias RL parameters $\theta$ via Q-learning in order to tackle the inaccurate above MPC model. In this example, the following optimization problem as a parameterized MPC scheme is solved at each time instant $k$:
    \begin{subequations}
		\begin{align}
			\min_{\vect x,\vect u,\vect \sigma }\quad &\theta_c+\frac{\gamma^{N^f}}{2}\left(\vect x^\top_{N^f}M^f_\theta \vect x_{N^f}+\vect w^\top_f\vect\sigma_{N^f}\right) \nonumber\\
			&+\sum_{i=k}^{k+N_{\text{MPC}}-1}\frac{\gamma^{i}}{2}\left((c-\theta_r)^\top M_\theta(c-\theta_r)+\vect w^\top\vect \sigma_{i}\right)\label{eq:v12}\\
			\mathrm{s.t.}&\quad \vect x_{i+1}=f_{\theta}^{\text{MPC}}(\vect x_i,\vect u_i),\label{eq:mpc_bias2}\\
			&\quad\vect x_k=\hat{\vect x}_k, \label{eq:v22}\\
			&\quad-1\leq\vect u_i\leq1,\\
			&\quad\begin{bmatrix}
                   0\\ 
                   -10
                  \end{bmatrix}+\underline{\theta}-\vect\sigma_i\leq\vect x_i\leq 
                  \begin{bmatrix}
                   10\\ 
                   10
                  \end{bmatrix}+\bar{\theta}+\vect\sigma_i\\
			&\quad \vect\sigma_{k,\ldots,k+N_{\text{MPC}}-1} \geq 0 \label{eq:violation2}
		\end{align}
	\end{subequations}
	where $N^f=k+N_{\text{MPC}}$, $c=\left[\vect x_k,\vect u_k\right]^\top$ and the MPC parameters subject to the RL scheme are:
	\begin{align}
	    \theta=\left(\theta_c,M^f_\theta,\theta_r,M_\theta,A^{\text{bias}},B^{\text{bias}},\underline{\theta},\bar{\theta}\right)
	\end{align}
	The positive semidefinite weighting matrices ($M^f_\theta$, $M_\theta$,$Q^\theta_E$, $R^\theta_E$, $A^\theta_{r}$) in both MPC and MHE schemes are adjusted using the constrained RL steps in \eqref{eq:sdp}.
	One can choose a baseline stage cost used in the RL scheme \eqref{eq:QL} as:
	\begin{gather}\label{eq:BL}
		L(\vect x_k,\vect u_k)=l(\vect x_k,\vect u_k)+\vect w^\top \max(0,\vect h(\vect x_k))
	\end{gather}
	 where $l(\vect x_k,\vect u_k)$ is adopted as a quadratic function of the state and action deviations from their desired values. The second term in the above baseline is considered to penalize the constraint violations, where $\vect h\geq 0$ is pure inequality vector of constraints on the states and $\vect w=[10,10]$. Note that different step sizes $\alpha$ were used for the different parameters based on the problem scaling. The desired values for the MPC model states (position and velocity of the first body) $x_1,x_3$ are chosen at $[0,0]^T$ , respectively. We apply a process noise $\zeta\sim \mathcal{N}(\mu=0,\,(\sigma=0.02)^{2})$ on the velocity of the second body and a measurement noise $\eta\sim \mathcal{N}(\mu=0,\,(\sigma=0.05)^{2})$ on the position of second body. In this simulation we choose $\gamma=0.9$, and $N_{\text{MPC}}=N_{\text{MHE}}=8$. 
	 \par
\begin{figure}[htbp!]
		\centering
		\includegraphics[width=0.9\linewidth]{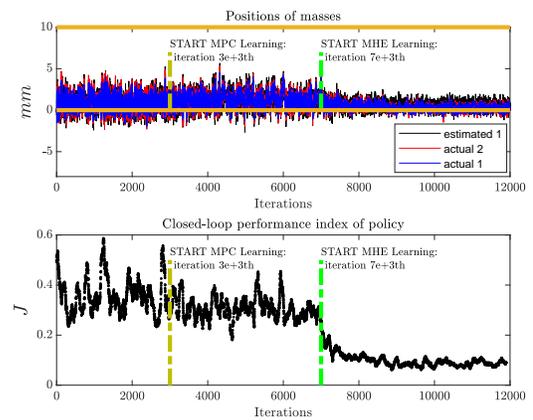}
		\caption{Positions of masses and closed-loop performance. The brown lines are shown as lower bound (0 $mm$) and upper bound (10 $mm$) constraints on the positions. Position references are (0 $mm$).}
		\label{modell}
	\end{figure}

 As this simulation is considered as a POMDP and uncertain scenario and there are both process and measurement Gaussian noises, the violations are observed on the states in Figure~\ref{modell}. We demonstrate that the proposed MPC/MHE-based RL can attenuate these violations and increase the control performance even if the controller/observer models are unmodeled and partially observable.
	\par In this example, we consider three different scenarios.  
	\par
\begin{figure}[htbp!]
		\centering
		\includegraphics[width=0.9\linewidth]{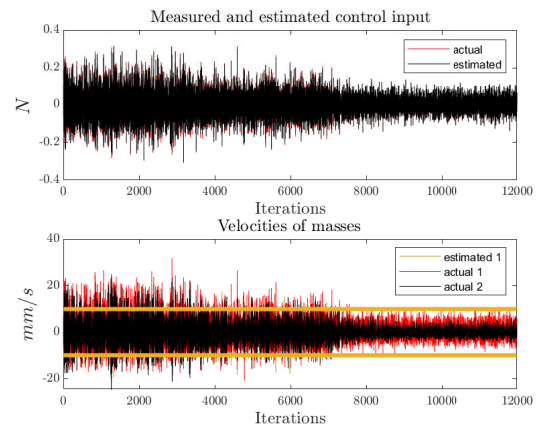}
		\caption{Control input and velocities of masses. The brown lines are shown as lower bound (-10 $mm/s$) and upper bound (10 $mm/s$) constraints on the velocities.}
		\label{Cont_Vel}
	\end{figure}

 1) \textit{Without learning} (MPC+MHE): In the first scenario, there are large violations of the position and velocity constraints affecting the closed-loop performance (large cost $J$) shown in Figure~\ref{modell} and Figure~\ref{Cont_Vel}. 
	\par
\begin{figure}[htbp!]
		\centering
		\includegraphics[width=0.9\linewidth]{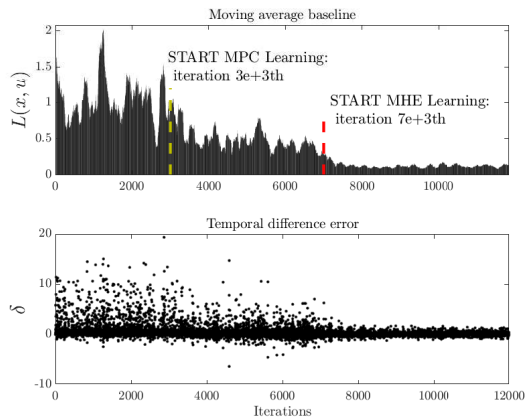}
		\caption{RL performance: baseline cost and TD error}
		\label{RL_BL_TD}
	\end{figure}

 2) \textit{MPC-based RL learning} (MPC-RL+MHE): In the second scenario, the learning is only performed on the MPC scheme. This MPC-based RL reduces the violations and increases the closed-loop performance shown in Figure~\ref{modell} by reducing the discounted sum of the RL stage cost $J$ over a receding horizon. There is also a decrease of the baseline cost shown in Figure~\ref{RL_BL_TD} after starting the MPC learning while the MHE learning is not still activated. 
	\par
\begin{figure}[htbp!]
		\centering
		\includegraphics[width=.9\linewidth]{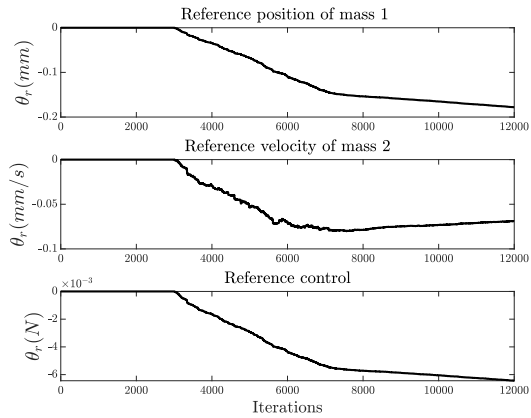}
		\caption{MPC adjustment: reference signals}
		\label{RL_ref}
	\end{figure}

\begin{figure}[htbp!]
		\centering
		\includegraphics[width=0.9\linewidth]{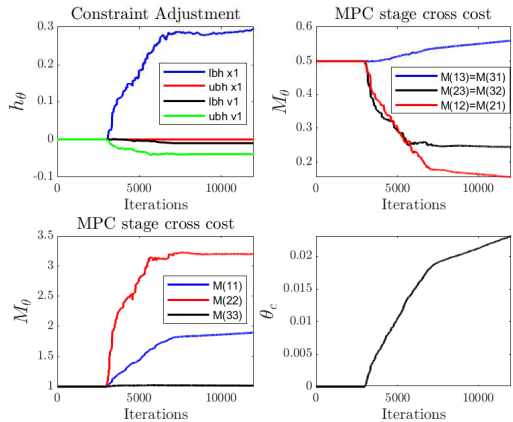}
		\caption{MPC adjustment: Constraint and stage cost}
		\label{RL_stage}
	\end{figure}

\begin{figure}[htbp!]
		\centering
		\includegraphics[width=0.9\linewidth]{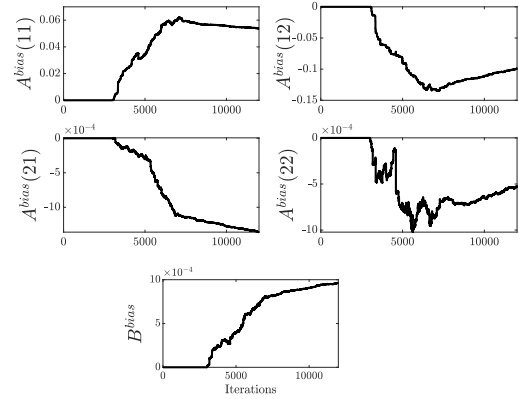}
		\caption{MPC adjustment: model bias}
		\label{RL_bias}
	\end{figure}

\begin{figure}[htbp!]
		\centering
		\includegraphics[width=0.8\linewidth]{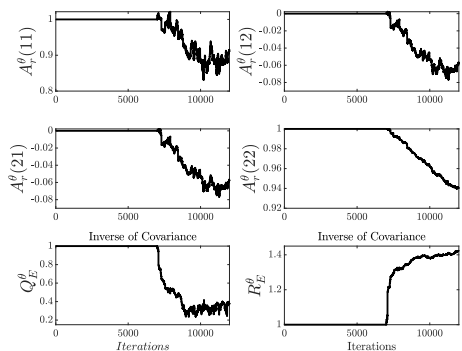}
		\caption{MHE adjustment: Arrival matrix and penalizing weights}
		\label{RL_arv}
	\end{figure}

\begin{figure}[htbp!]
		\centering
		\includegraphics[width=0.8\linewidth]{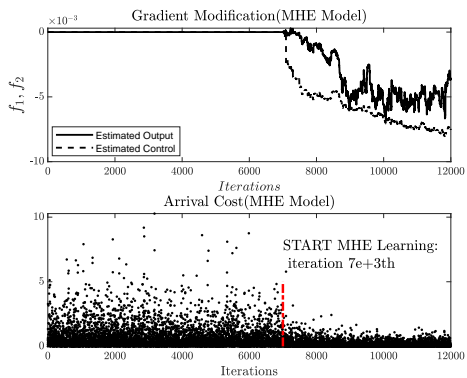}
		\caption{Evolution of arrival cost and gradients in MHE}
		\label{cost_arv}
	\end{figure}

 3) \textit{MPC/MHE-based RL learning} (MPC-RL+MHE-RL): Finally in the third scenario, the performance is improved after allowing the MHE to be adjusted using the Q-learning algorithm and there is a solid decrease in the TD-error and baseline cost and an increasing closed-loop performance (decrease of $J$). The evolution of MHE parameters are illustrated in Figure~\ref{RL_arv} and Figure~(\ref{cost_arv}). The evolution of MPC parameters are depicted in Figures \ref{RL_ref}, \ref{RL_stage}, and \ref{RL_bias}. 
	\par 
	\section{Conclusion} \label{sec:Conclusion}
	This paper proposed the combination of MPC-based Reinforcement Learning with an MHE scheme to tackle POMDPs. The introduction of an MHE scheme allows to deploy MPC-based Reinforcement Learning without a full state measurement, and without necessarily holding a correct representation of the system state in the MPC and MHE models. Furthermore, we propose to tune the MHE and MPC schemes jointly, focusing directly on the closed-loop performance, as opposed to using indirect criteria such as decreasing the MHE output error. We detail the application of Q-learning to this approach, and test it in a simulated spring mass example operating under constraints, where only a part of the real system dynamics are modelled in the MPC and MHE schemes. We show that the method manages to tune the MHE and MPC scheme to reduce the constraints violations and improve the closed-loop performance. Future work will propose an stability and feasibility analysis on the proposed MPC/MHE-based RL scheme.
	\bibliographystyle{IEEEtran}
	\bibliography{IEEEabrv,main}
\end{document}